\documentclass[sort&compress,numbers]{aastex63}

\usepackage{graphics,graphicx}
\usepackage{hyperref}

\newcommand{\skipthis}[1]{}
\newcommand{\hii}{{\rm H}{\sc ii}}

\def\kms{km~s$^{-1}$}
\def\msun{$M_\odot$}

\def\mjyb{mJy beam$^{-1}$}

\def\nh3{$\rm{NH_3}$}
\def\NH3{$\rm{NH_3}$}

\def\h2{$\rm{H_2}$}

\def\cm2{$\rm{cm^{-2}}$}
\def\cm3{$\rm{cm^{-3}}$}

\def\hii{H\,{\sc ii}}

\renewcommand{\arraystretch}{1.2}

\accepted{to ApJ Letters, December 7, 2022}

\shorttitle{Clustered Formation of Massive Stars within an Ionized Rotating Disk}
\shortauthors{Galv\'an-Madrid et al.}

\graphicspath{{./}}

\begin{document}

\title{Clustered Formation of Massive Stars within an Ionized Rotating Disk}


\author[0000-0003-1480-4643]{Roberto Galv\'an-Madrid}
\affil{Instituto de Radioastronom\'ia y Astrof\'isica, Universidad Nacional Aut\'onoma de M\'exico, Morelia, Michoac\'an 58089, M\'exico.}

\author[0000-0003-2384-6589]{Qizhou Zhang}
\affil{Center for Astrophysics \textbar\, Harvard \& Smithsonian, 60 Garden St., Cambridge, MA 02138, USA.}

\author[0000-0001-8446-3026]{Andr\'es Izquierdo}
\affil{European Southern Observatory, Karl-Schwarzschild-Str. 2, D-85748 Garching bei M\"unchen, Germany.}
\affil{Leiden Observatory, Leiden University, P.O. Box 9513, NL-2300 RA Leiden, The Netherlands.}

\author[0000-0003-1413-1776]{Charles J. Law}
\affil{Center for Astrophysics \textbar\, Harvard \& Smithsonian, 60 Garden St., Cambridge, MA 02138, USA.}

\author{Thomas Peters}
\affil{Max-Planck-Institut f\"{u}r Astrophysik, Karl-Schwarzschild-Str. 1, D-85748 Garching, Germany.}

\author{Eric Keto}
\affil{Center for Astrophysics \textbar\, Harvard \& Smithsonian, 60 Garden St., Cambridge, MA 02138, USA.}

\author[0000-0003-2300-2626]{Hauyu Baobab Liu}
\affil{Physics Department, National Sun Yat-Sen University, No. 70, Lien-Hai Road, Kaosiung City 80424, Taiwan, R.O.C.}
\affil{Institute of Astronomy and Astrophysics, Academia Sinica, 11F of Astronomy-Mathematics Building, AS/NTU No.1, Sec. 4, Roosevelt Rd, Taipei 10617, Taiwan, ROC.}

\author[0000-0002-3412-4306]{Paul T. P. Ho}
\affil{Institute of Astronomy and Astrophysics, Academia Sinica, 11F of Astronomy-Mathematics Building, AS/NTU No.1, Sec. 4, Roosevelt Rd, Taipei 10617, Taiwan, ROC.}
\affil{East Asian Observatory (EAO), 660 N. Aohoku Place, University Park, Hilo, Hawaii
96720, USA.} 

\author[0000-0001-6431-9633]{Adam Ginsburg}
\affiliation{Department of Astronomy, University of Florida, P.O. Box 112055, Gainesville, FL, USA.}

\author[0000-0003-2862-5363]{Carlos Carrasco-Gonz\'alez}
\affil{Instituto de Radioastronom\'ia y Astrof\'isica, Universidad Nacional Aut\'onoma de M\'exico, Morelia, Michoac\'an 58089, M\'exico.}

\begin{abstract}
We present ALMA observations with a 800 au resolution and radiative-transfer modelling of the inner part ($r\approx6000$ au) of the ionized accretion flow around a compact star cluster in formation at the center of the luminous ultra-compact (UC) \textsc{Hii} region G10.6-0.4. We modeled the flow with an ionized Keplerian disk with and without radial motions in its outer part, or with an external Ulrich envelope. The MCMC fits to the data give total stellar masses $M_\star$ from 120 to $200~M_\odot$, with much smaller ionized-gas masses $M_\mathrm{ion-gas} = 0.2$ to $0.25~M_\odot$. The stellar mass is distributed within the gravitational radius $R_g\approx 1000$ to 1500 au, where the ionized gas is bound. The viewing inclination angle from the face-on orientation is $i = 49$ to $56~\deg$. Radial motions at radii $r > R_g$ converge to $v_{r,0} \approx 8.7$ \kms, or about the speed of sound of ionized gas, indicating that this gas is marginally unbound at most. 
From additional constraints on the ionizing-photon rate and far-IR luminosity of the region, we conclude that the stellar cluster consists of a few massive stars with $M_\mathrm{star} = 32$ to $60~M_\odot$, or one star in this range of masses accompanied by a population of lower-mass stars.  
Any active accretion of ionized gas onto the massive (proto)stars is residual. 
The inferred cluster density is very large, comparable to that reported at similar scales in the Galactic Center. Stellar interactions are likely to occur within the next Myr.  
\end{abstract}


\section{Introduction} \label{sec:intro}
Stars with masses greater than 30 \msun~are rare among the stellar population in galaxies, but they play a main role in shaping the dynamical evolution of the interstellar medium through feedback from their radiation, ionization, and supernovae \citep[for a review of feedback effects in star formation, see][]{krumholz2014}. Since these very massive stars are mostly formed in dense stellar clusters, their violent death may lead to the formation of black holes or neutron-star binary systems that give rise to gravitational-wave events \citep[e.g.,][]{giacobbo2018}. 
Massive protostars must accrete more than 30~\msun~to become early O-type stars, but they begin core hydrogen burning well before reaching their final masses \citep{palla1993,hosokawa10}. This poses a serious challenge to the current paradigm that massive stars form through accretion of molecular gas. The detection of resolved molecular disks around protostars more massive than $20~M_\odot$
shows that this scenario is valid for the formation of individual stars up to about a few tens of solar masses \citep[e.g.,][]{johnston2015,ilee2018,sanna2019,olguin2020,lu2022}; however, it likely breaks down for the case of the formation of more massive stars, especially in clustered environments  \citep{goddi2020}. In a clustered scenario, and after the protostars have reached several tens of solar masses, the amount of EUV photons is large enough to ionize the otherwise molecular accretion flow(s) and the clump in which they reside \citep{keto2003,keto2007,peters2010b,peters2010a,klaassen2018}. This raises several outstanding questions: How does star formation proceed in such environment? Is it through an ionized accretion disk? Does this disk surround a single or a multiple stellar system? When does active (proto)stellar accretion stop due to the increasing photoionizing feedback? 

We present the first spatially resolved observations of an  accretion flow in ionized gas around a dense star cluster in formation. Our observations were performed with the Atacama Large Millimeter/submillimeter Array (ALMA) in the 1.3 mm continuuum and hydrogen $30\alpha$ recombination line emission. 
The target is G10.6-0.4 (hereafter G10.6), a  luminous ($L_\mathrm{FIR} \approx 3\times10^6~L_\odot$) and spatially concentrated \citep[see][]{lin2016} massive star formation region located at 4.95 kpc from the Sun \citep{sanna2014}. The molecular gas in the central pc-scale clump of G10.6 presents infall and rotation motions toward an inner, flattened ultracompact \textsc{Hii}~region \citep[e.g.,][]{ho1986,keto1987a,sollins2005b,liu2011,beltran2011}.  A characteristic ``bullseye'' velocity pattern in the most redshifted molecular absorption against the central UC \textsc{Hii}~region reveals that the molecular infall is coupled with rotation \citep[see][]{sollins2005b}. By equating the observed molecular infall speed to free fall, \citet{sollins2005b} estimated a central stellar mass of $M_\star \sim 150$. \citet{keto2006} reported a velocity gradient in the H$66\alpha$ recombination line within the central 0.05 pc. 
Those authors proposed that the central ionized gas is an accretion flow feeding a nascent cluster of massive stars.

\section{Observations} \label{sec:obs}

Observations of G10.6 with ALMA (project ID: 2015.1.00106.S, PI Q. Zhang) were carried out in four execution blocks (EBs) between 2016 and 2017. The total on-source integration time was 2.2 hours. The two sessions executed on 2016 September 9 and 10 had 36 12-m antennas in the array with physical baselines from 15 m to 3144 m. The remaining sessions were observed in 2017 July 19 had 42 12-m antennas with baselines from 19 m to 3697 m. The precipitable water vapor (PWV) in the atmosphere ranged between 0.46 to 0.50 mm during the observations. System temperatures varied from 49 to 75 K. Quasars J1832-2039, J1924-2914 and J1733-1304 were used as gain, bandpass, and flux calibrators, respectively. 

The digital correlator was configured to the FDM mode for four spectral windows, centered at 217.9~GHz, 220.0~GHz, 232.0~GHz and 233.9~GHz, respectively. Each spectral window has an effective bandwidth of 1.875~GHz, divided into 1910 channels, providing a uniform spacing of 0.976~MHz per spectral channel. The pointing center of G10.6 was $\alpha$(J2000) = 18h 10m 28.7s, $\delta$(J2000) = $-19^\circ 55' 49.1''$.

The calibration of the visibilities was performed in Common Astronomy Software Applications package \citep[\texttt{CASA}][]{mcmullin2007} using the pipeline script supplied by the ALMA observatory. The calibrated visibilities were Fourier transformed and `cleaned' using {\it tclean} and Briggs weighting of visibilities with robust parameter of 0.5. Spectral channels free of significant line emission are used to construct continuum data. Continuum emission was subtracted from the calibrated visibilities to produce the spectral line visibilities. The rms noise in the continuum image is about 35 $\mu$Jy~beam$^{-1}$.  The rms noise in the line free channels of the H30$\alpha$ image cube is 0.5 mJy~beam$^{-1}$ per 1.5 \kms~channel, with a beam of $0''.15 \times 0''.13$, $\mathrm{P.A.} = -65.8 \deg$. 

\section{Results} \label{sec:results}

\subsection{Overview} \label{sec:overview}

\begin{figure*}
\hspace{-2.0cm}
\begin{center}
\includegraphics[width=\textwidth]{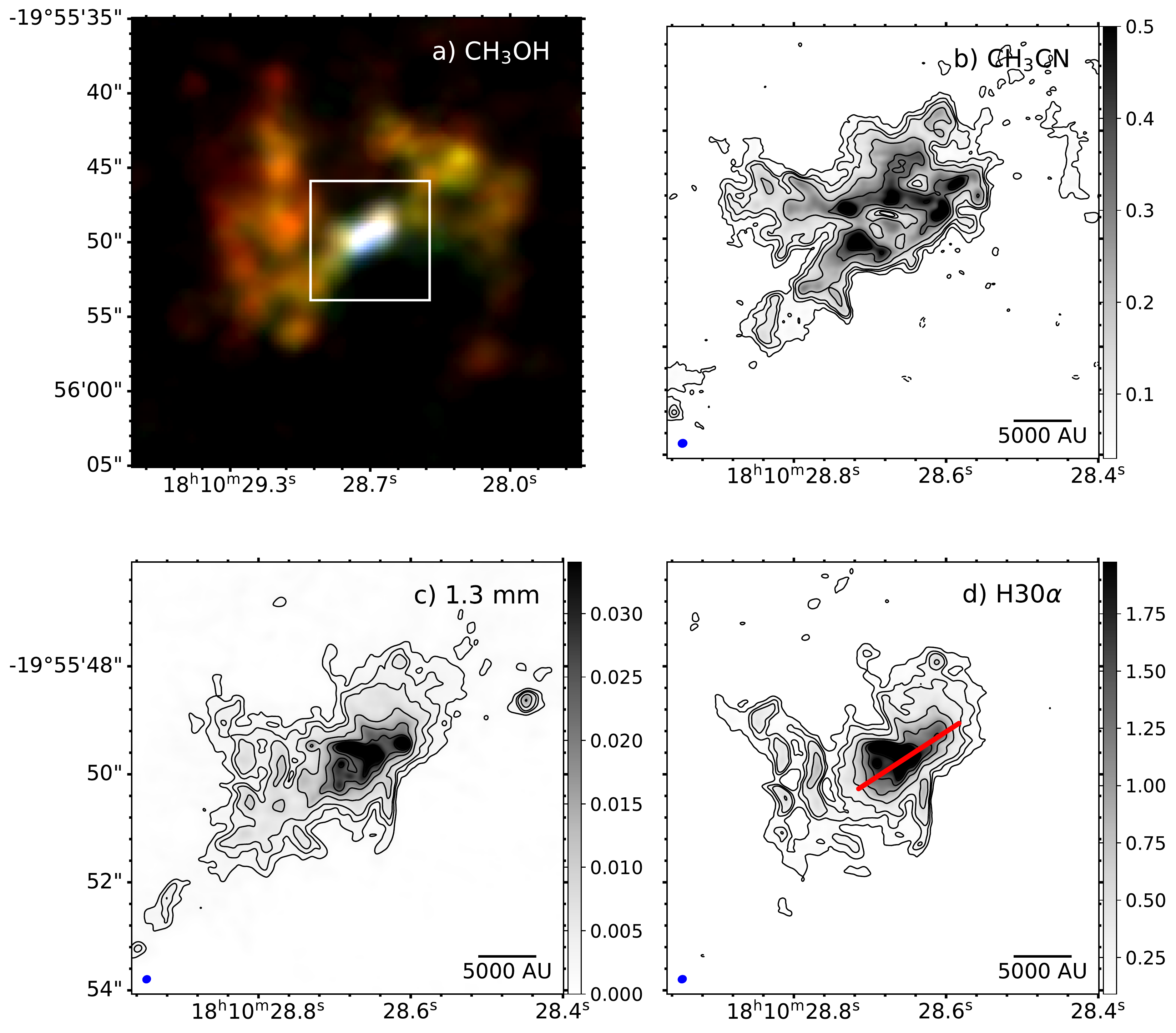}
\end{center}
\caption{Distribution of molecular and ionized gas in the dense cluster G10.6-0.4. {\it (a):} CH$_3$OH line emission imaged with the SMA. Red, green, and blue show CH$_3$OH with upper energies 35, 59, and 97 K \citep[see][]{liu2011}. The white square marks the region of the ALMA maps shown in the other panels. {\it (b):} Emission of the CH$_3$CN J=12-11 line~\citep[see][]{law21} [Jy beam$^{-1}$ km s$^{-1}$].  
{\it (c):} 1.3-mm continuum [Jy beam$^{-1}$].  
{\it (d):} Integrated H30$\alpha$ line emission [Jy beam$^{-1}$ km s$^{-1}$]. The spatial cut for the P-V diagrams is shown with a red line. 
}
\label{fig:Fig1}
\end{figure*}

The ALMA observations with a physical resolution of 792 au ($0.16 \arcsec$ at a distance of 4950 pc) resolve the 1.3 mm continuum and H${30}\alpha$ emission within the inner 6000 au radius of the star-forming clump. 
Figure \ref{fig:Fig1} presents an overview. 
The top-left panel shows CH$_3$OH emission imaged with the Submillimeter Array. The molecular envelope of a $\sim 0.3$ pc radius flattens toward its center, most notably in the warmer (blue) CH$_3$OH transition \citep[see][]{liu2011}. An evacuated bipolar cavity is also seen. 
The emission from complex organic molecules (COMs) in the inner 0.05 pc was recently studied by \citet{law21} using ALMA. Those authors found highly structured emission in the form of localized hot cores accompanied by bright, extended emission. The overall COM emission presents a flattened X-shaped morphology with a brightness dip toward the H${30}\alpha$ peak (see Fig. \ref{fig:Fig1}, CH$_3$CN panel). In contrast, the 1.3-mm continuum peaks at the central dip of the COM emission and matches well the H${30}\alpha$ emission (Fig. \ref{fig:Fig1}, bottom panels). 
The continuum emission is almost entirely due to the free-free radiation in the region that we analyze. The average continuum intensity within a stripe of $1\arcsec$ length and 1 beamwidth height is 0.045 Jy beam$^{-1}$ (40 K), whereas the free-free continuum expected from the H30$\alpha$ intensity in the same area, using equation (14.29) in \citet{rohlfs2000}, is 0.044 Jy beam$^{-1}$.
Contributions from warm dust to the continuum become important at the location of hot cores in the periphery of the central ionized structure \citep{liu2010b,law21}. This flattened ionized emission within 0.05 pc radius has been interpreted as the innermost part of a cluster accretion flow which transitions from being molecular to ionized at its center \citep[e.g.,][]{keto1987a,keto2006}. 

The LSR systemic velocity of the H$30\alpha$ line is determined to be $V_\mathrm{sys} \approx 0.5$ \kms. The H$30\alpha$ panel of Fig. \ref{fig:Fig1} shows the directional cut at $\mathrm{PA} = 307.8^\circ$ (East of North) used for the modelling. Figure \ref{fig:Fig2} shows the resulting position-velocity (P-V) diagram. High-velocity emission characteristic of Keplerian rotation 
is seen on either side from the position center, reaching velocities up to $\pm \sim 40$ \kms~from $V_\mathrm{sys}$ in the inner few $10^3$~au. 
This kinematic structure is in direct contrast to that of classical \hii~regions, in which the photoionized gas at $10^4$ K exerts an outward pressure that drives an unimpeded expansion. 
We rather interpret our observations in the context of \hii ~regions still dominated by the gravity of the embedded stars \citep{keto2002b, keto2003,keto2007}. 
In these models the ionized gas within the gravitational radius $ R_g = GM_*/c_s $ (where $c_s$ is the sound speed of the ionized gas) remains bound, and can continue to accrete onto the stars. 

\subsection{Radiative Transfer and Model Fitting} \label{sec:modelling}

We use \texttt{RADMC-3D}  \citep{dullemond12} to model the 1.3-mm free-free continuum and the H$30\alpha$ emission in non-LTE. We use the version that allows to calculate recombination lines as  presented in \citet{peters2012}. The physical 3D model grids are created with \texttt{sf3dmodels} \citep{izquierdo18} and manipulated with the tools within that package to be compatible with \texttt{RADMC-3D}. 

The observational data sets that we model are a 1D intensity cut of the 1.3-mm continuum and the 2D P-V diagram of the H${30}\alpha$ emission. Both cuts are centered at ICRS RA $=$ 18h 10m 28.652s,  Dec $=$ $-19$d 55m 49.66s, $\mathrm{PA} = 307.8^\circ$. The length of the P-V and continuum cuts are 2.32\arcsec (11480 au) and 0.72\arcsec (3560 au), respectively, and their width is 0.16\arcsec.  The continuum cut was truncated because the emission deviates from a single power law beyond the defined length. This is probably due to a real change in the density profile of the ionized disk, or due to contributions from dust emission at larger radii (see Section \ref{sec:overview}). The continuum intensity cut is only used in the model fitting to constrain the distribution of the electron density $n_e$. We consider fully ionized hydrogen gas ($n_e = n_{ion}$). The kinematic structure is as follows.

\textit{A Keplerian rotating disk.} The simplest model is a flared Keplerian disk as in \citet{pringle1981}. 
The disk electron density $n_e$ goes as: 
\begin{equation} \label{eq:disk-dens}
n_e(R,z)= n_0 \biggl [ \frac{R}{R_0} \biggr ]^p \exp [-z^2 / 2H^2], 
\end{equation} 

\noindent
where $R$ is the polar radius normalized at $R_0 = 10$ au, $z$ is the distance from the midplane, and $H(R) \propto R$ is the scale height. For simplicity, we fix $H$ to have a linear dependency with radius, but we have verified that steeper radial variations $H(R) \propto R^{1.25}$ or $\propto R^{1.5}$ only affect the results minimally. The normalization of $H$ is selected such that at the gravitational radius $H(R_g) = R_g$, in accordance with the physical expectation that within $R_g$ the ionized gas is bound \citep{hollenbach1994,keto2007,tanaka2013}. These assumptions leave only two free parameters for the density description in Equation \ref{eq:disk-dens}: its normalization $n_0 = n_e(R=10~\mathrm{au})$, and its  power-law index $p$. 

\smallskip
The circular velocity in the Keplerian model is defined as: 
\begin{equation} 
v(R) = \biggl[ \frac{GM_\star}{R} \biggr ]^{0.5}, 
\end{equation}

\noindent
where $G$ is the gravitational constant and $M_\star$ is the central stellar mass. This kinematical model has two free parameters: $M_\star$ and the inclination angle $i$, where $i=90 \deg$ means an edge-on view. 

\smallskip
\textit{A Keplerian disk with outer radial motions.} The next level of complexity is to include radial spherical motions to the disk, which are vector-added to the Keplerian rotating disk. 
This adds a third kinematical free parameter $v_{r,0}$.

\smallskip
\textit{A Keplerian disk with an Ulrich envelope.} 
We also considered a model consisting of a Keplerian disk surrounded by an Ulrich-type envelope \citep{ulrich1976,mendoza04}. This model has been previously used to interpret the kinematics of G10.6 at $\sim 0.1$ pc scales \citep{keto2006}, and is a widely-used option to interpret rotating-infalling envelopes which settle into a rotationally supported disk \citep[e.g.,][]{keto2010,izquierdo18}. 
By construction, the radial motions in the outer part of this model are inward.  
We run models with four free parameters: the stellar mass $M_\star$, the viewing inclination angle $i$, the accretion rate of the Ulrich envelope $\dot{M}_\mathrm{env}$, and an adimensional scaling factor $A$ which controls the density contrast between the inner disk and the outer envelope. 

\begin{deluxetable*}{ccccccccccc}[t]
\tablenum{1}
\scriptsize
\tablewidth{0pt}
\tablecaption{Model Fitting \label{tab:model-fitting}}
\tablehead{
\multicolumn{2}{l}{Density profile} & 
\multicolumn{2}{l}{Keplerian disk} & 
\multicolumn{3}{c}{Keplerian $+$ radial} & 
\multicolumn{4}{c}{Keplerian $+$ Ulrich}
\\ 
\hline 
\colhead{$n_0$} & \colhead{$p$} & \colhead{$M_\star$} & \colhead{$i$} & 
\colhead{$M_\star$} & \colhead{$i$} & \colhead{$v_\mathrm{r,0}$} & 
\colhead{$M_\star$} & \colhead{$i$} & \colhead{$\dot{M}_\mathrm{env}$} & \colhead{$A$} \\
$[10^{12}~\mathrm{cm}^{-3}]$ & & $[\mathrm{M_\odot}]$ & $[\deg]$ & $[\mathrm{M_\odot}]$ & $[\deg]$ & [\kms] & $[\mathrm{M_\odot}]$ & $[\deg]$ & $[10^{-5}~\mathrm{M_\odot~yr}^{-1}]$ & 
}
\startdata
$5.5\pm0.1$ & $-0.67\pm0.01$ & $194.0\pm0.8$ & $51.2\pm0.1$ & 
$127.9\pm0.8$ & $55.5\pm0.1$ & $8.72\pm0.04$  &
$187.8\pm0.9$ & $48.9\pm0.2$ & $3.33\pm0.02$ & $5.29\pm0.03$  
\enddata 
\tablecomments{Continuum fit for density profile parameters: density normalization $n_0$ and index $p$. H$30\alpha$ fit for kinematical parameters: Keplerian model -- stellar mass $M_\star$, viewing inclination angle $i$; Keplerian model with external radial motions -- $M_\star$, $i$, and radial velocity $v_\mathrm{r,0}$; Keplerian model with external Ulrich envelope -- $M_\star$, $i$, accretion rate of the envelope $\dot{M}_\mathrm{env}$, disk density scaling factor $A$. 
The statistical errors ($\pm 1\sigma$) from the MCMC fitting are defined to contain $68.2~\%$ of the values. 
}
\end{deluxetable*}


Table \ref{tab:model-fitting}  summarizes the free parameters of the different modelling scenarios. Figures \ref{fig:App1-Fig1}, \ref{fig:App1-Fig2}, and \ref{fig:App1-Fig3} in the Appendix show diagnostic plots for each of the best-fit models. 
The modelling and fitting procedure is as follows: 
\begin{itemize}
    \item Homogenize the continuum and line data to a common circular beam of $\mathrm{FWHM} = 0.16\arcsec$ (792 au), with a pixel size of $0.08\arcsec$. Nyquist sampling was chosen to avoid overfitting the data.  
    \item Extract the observational continuum intensity and P-V cuts as previously described. 
    \item Create a 3D model with \texttt{sf3dmodels} \citep{izquierdo18} and export it to the grid format of \texttt{RADMC-3D} \citep{dullemond12,peters2012}.  
    \item Use \texttt{RADMC-3D} to calculate the radiative transfer of the free-free continuum and non-LTE H$30\alpha$ emission of the previously computed model. 
    \item Convolve the model images to the same resolution and pixel size of the observations. Then extract the model continuum and P-V cuts.  
    \item Compare observations and model, iterating over the model parameter space using the MCMC sampler implemented in \texttt{emcee} \citep{foreman13}. 
\end{itemize}

The log-likelihood function that is maximized is defined as: 
\begin{equation} \label{eq:logchi2}
\chi^2 = -0.5 \sum \biggl ( \frac{I_\mathrm{data} - I_\mathrm{model}}{\sigma_\mathrm{noise}} \biggr )^2,
\end{equation}     

\noindent
where the summation is over the 9 pixels of the continuum cut or the $29\times60=1740$ pixels of the H${30}\alpha$ P-V image, depending on the respective maximization\footnote{Note that the number of independent measurements is half of the number of pixels. We have verified that changing the number of pixels per beam does not affect the results.}. 
To get a conservative estimate of the noise in the fitted data, for the line modelling we measure the rms in the P-V diagram $\sigma_\mathrm{noise} = 1.5$ \mjyb. Note that it is larger than the rms noise measured in a single channel in a region free of emission (see Section \ref{sec:obs}), since it contains contributions from channel-to-channel variations and correlated noise due to interferometric sidelobes. For the continuum modelling we set $\sigma_\mathrm{noise} = 0.5$ \mjyb. 

\begin{figure*}
\hspace{-2.0cm}
\centering
\includegraphics[width=\textwidth]{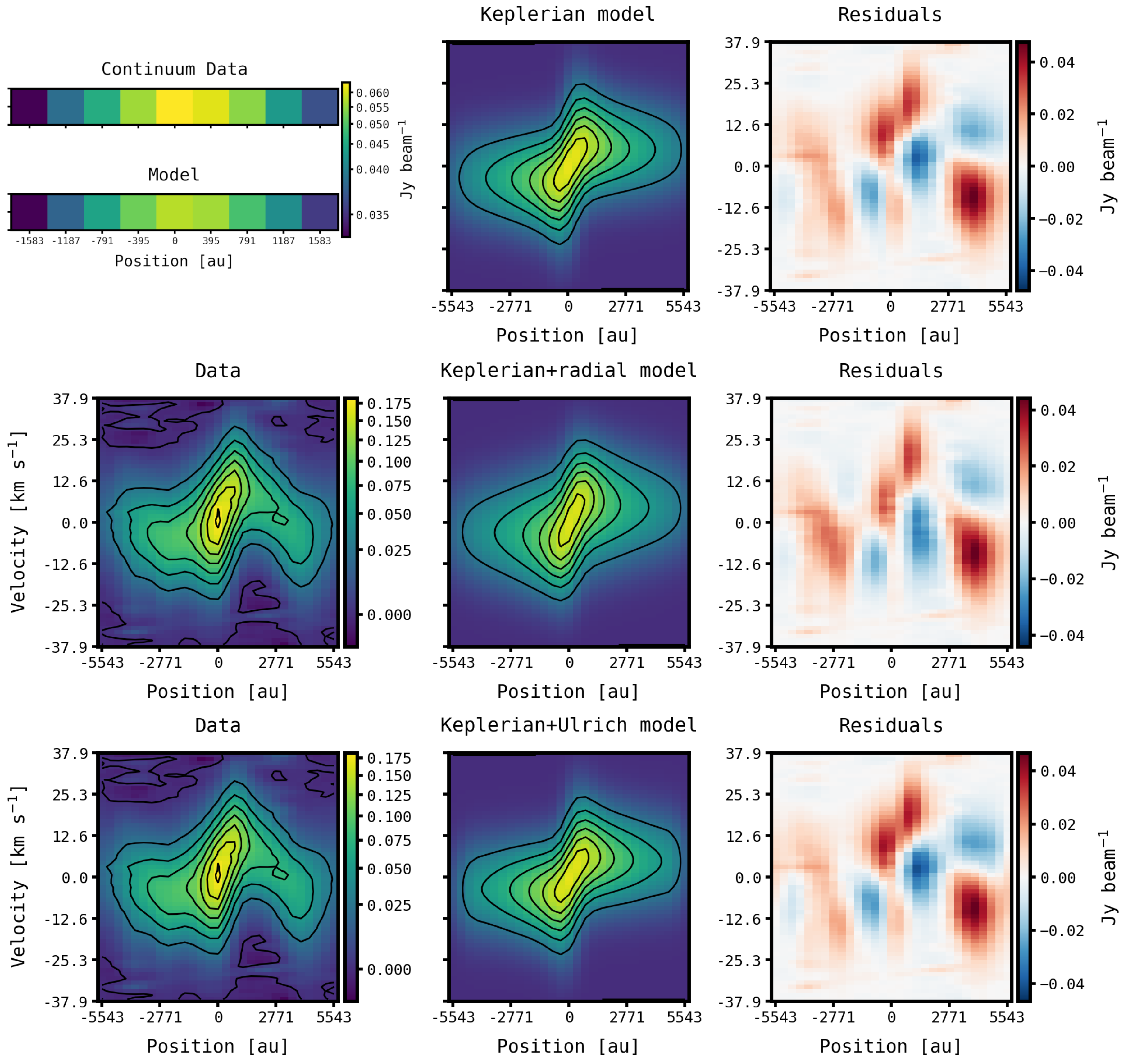}
\caption{Results from the radiative-transfer model fitting. 
{\it Top row, left panel}: continuum modelling. The top cut shows the ALMA data and the bottom cut shows the best-fit power law model for the electron density. 
{\it Top row, center and right panels}: Keplerian disk best-fit and residuals.
{\it Middle row}: ALMA  H$30\alpha$ position-velocity diagram, along with the best fit Keplerian model with radial motions and residuals.
{\it Bottom row}: same as middle row for the Keplerian disk with Ulrich envelope. 
Residuals are defined as observational data minus model. 
}
\label{fig:Fig2}
\end{figure*}

The continuum modelling is used to find the two free parameters of the disk density profile: $n_0 = 5.5\times10^{12}$ cm$^{-3}$, $p=-0.67\approx-2/3$ (see Table \ref{tab:model-fitting}). These two density parameters are then fixed in the three different kinematical models of the H$30\alpha$ line. Figure \ref{fig:Fig2} shows the results for the continuum and line modelling. The best-fit models give similar residuals because their overall properties are similar (see figures in the Appendix). Figure \ref{fig:Fig3} shows the posterior probability ``corner'' plots for the different types of models. 
The statistical errors are very small ($\sim 1~\%$) and should be considered as strict lower limits. Our consideration of different scenarios shows that the main source of uncertainty is in the model assumptions.

For the case of a purely Keplerian model, the MCMC runs converge to $[M_\star, i] = [194.0~M_\odot, 51.2~\deg]$ . For the scenario of a Keplerian disk with radial motions, we initially set the radius at which radial motions start to the gravitational radius of the purely Keplerian model: $r_0 = R_g(M_\star = 194~M_\odot) = 2320$ au. The resulting central mass is consistently smaller ($\approx 130~M_\odot$) when radial motions are included compared to the purely Keplerian case, therefore we updated the fixed value of $r_0$ to 1550 au, or about $R_g$ for a central mass of $130~M_\odot$. 
The resulting values for the free parameters of this fit are $[M_\star, i, v_{r,0}] = [127.9~M_\odot, 55.5 \deg, 8.72$ \kms] (see Table \ref{tab:model-fitting}). The residuals from this model are slightly improved because  radial motions help to produce a larger velocity spread at radii beyond $\sim 1000$ au (see Fig. \ref{fig:Fig2}).  
We have verified that this  result is insensitive to our selection of $r_0$. Including $r_0$ as a fourth free parameter gives consistent results: $[M_\star, i, v_{r,0}, r_0] = [123.3~M_\odot, 56.4~\deg, 8.0$ \kms$, 972$ au]. The assumption that radial motions start at the gravitational radius given by our simple prescription might be only approximately valid, since radiation pressure onto ions could decrease the radius at which outward radial motions start \citep{tanaka2017}.  
Interestingly, the results presented in Table \ref{tab:model-fitting} where  $v_{r,0}$ was restricted to be positive (i.e., outward radial motions) are identical if $v_{r,0}$ is restricted to be negative (inward motions). 
Therefore, the modelling cannot distinguish between these two scenarios because the H30$\alpha$ emission is optically thin. The arguments for the selection of $r_0 \sim R_g$ suggest an outward interpretation, but the infalling molecular gas in the exterior suggests otherwise (see Section \ref{sec:discussion}). It is also remarkable that the fitted radial motions $v_{r,0}$ are identical to the sound speed of ionized gas ($c_s = 8.6$ \kms) at the assumed electron temperature $T_e=9000$ K.  This means that, if unbound, the ionized gas in the disk has not yet accelerated to its terminal supersonic expansion \citep{franco1989}. Finally, the results for the kinematical models of a Keplerian disk with an external Ulrich envelope are $[M_\star, i, \dot{M}_\mathrm{env}, A] = [187.8~M_\odot, 48.9~\deg,3.3\times10^{-5}~M_\odot$ yr$^{-1}, 5.29]$ (see Table \ref{tab:model-fitting}). The obtained stellar mass and inclination angle are consistent with the values of the disk-only models. 

\begin{figure*}
\hspace{-2.0cm}
\centering
\includegraphics[width=\textwidth]{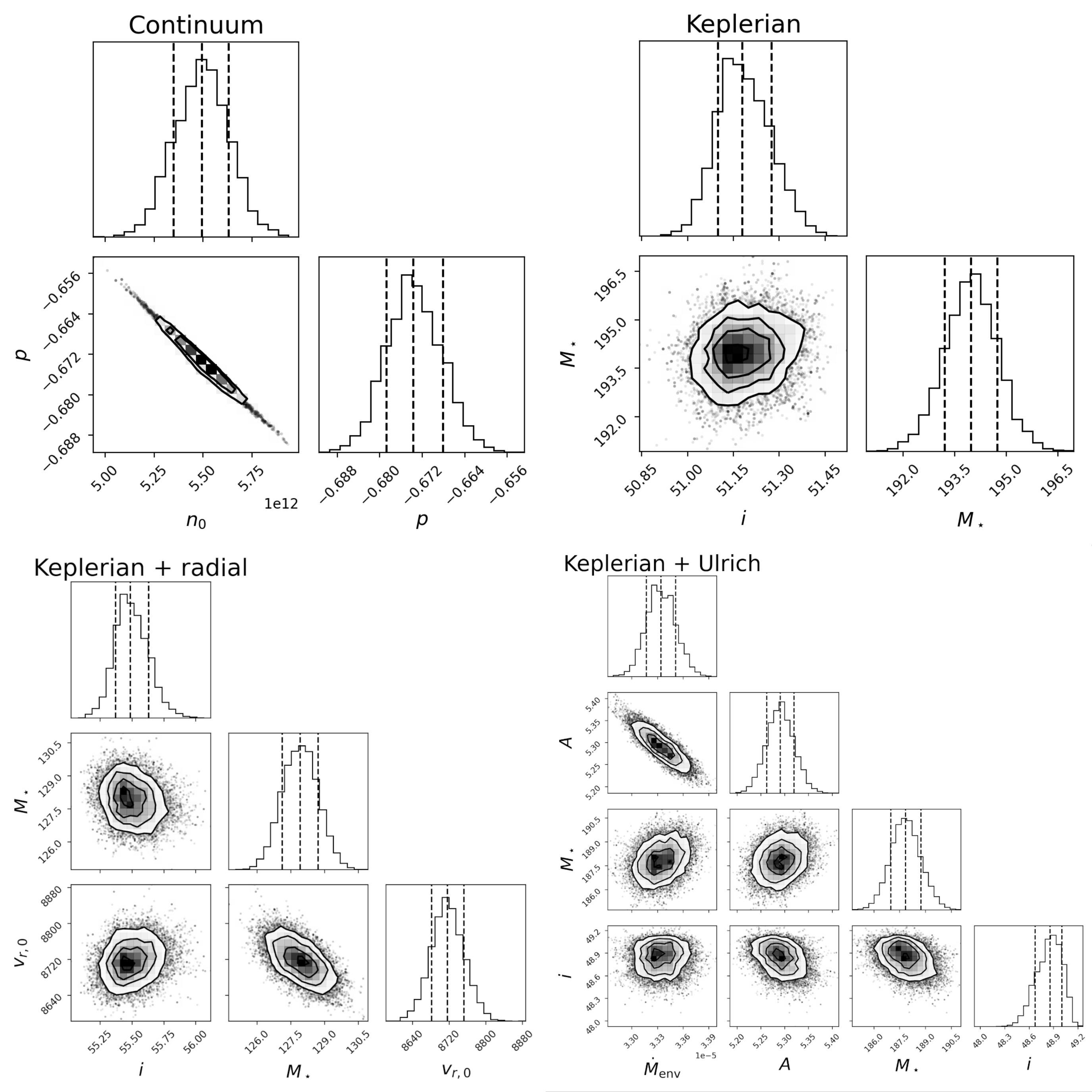}
\caption{Corner plots of posterior probability distributions of the model parameters fitted by maximizing the log-likelihood $\chi^2$ function defined in Equation \ref{eq:logchi2}. 
The {\it top-left} panels show the parameters labeled as ``Density profile'' in Table \ref{tab:model-fitting}. 
The {\it top-right} panels correspond to the ``Keplerian disk'' modelling. 
The {\it bottom-left} panels are for the ``Keplerian $+$ radial'' models. 
The {\it bottom-right} panels denote the ``Keplerian $+$ Ulrich'' models. 
}
\label{fig:Fig3}
\end{figure*}

\section{Discussion} \label{sec:discussion}

Our results make the strongest case to date for the central stellar mass of the G10.6 massive star forming clump to be very large, $M_\star = 120$ to $200~M_\odot$. 
The modelling results are consistent with the evidence presented by \citet{sollins2005b}, who used NH$_3$ inversion lines with a resolution similar to ours, but originating in the molecular infall zone exterior to the central ionized disk, to infer a central mass $M_\star \sim 150$ M$_\odot$. 
From our kinematical modelling, this mass is distributed within a radius $< 1000$ to 1500 au, since this is the part of the ionized structure that is dominated by Keplerian rotation. Moreover, the mass has to be (proto)stellar,  because the amount of ionized gas needed to reproduce the continuum and H$30\alpha$ brightness is relatively small. The best-fit models give a total amount of ionized gas of only 0.2 to $0.25~M_\odot$ within the entire domain of length 11480 au. 
Besides constraining the central mass of the cluster, three main questions remain: $(i)$ what is the distribution of (proto)stellar masses?; $(ii)$ what is the accretion stage of these (proto)stars?; and $(iii)$ what is the fate of this compact cluster? 

For question $(i)$,
the ionizing-photon rate $N_\mathrm{Ly}$ inferred from the free-free continuum gives independent constraints on the most massive stars. Our models give a range $N_\mathrm{Ly} = 6.1$ to  $6.6\times10^{48}$ s$^{-1}$, whereas the spherical estimation by \citet{sollins2005b} gives a significantly larger  $N_\mathrm{Ly} = 2\times10^{50}$ s$^{-1}$. The main reason for this  discrepancy is the rapid density decrease in the vertical direction in our models, which are tailored to fit the averaged midplane cuts of the flattened ionized structure. The UC \textsc{Hii} region is far from spherical\footnote{The spherical power law with a density index $n \propto r^{-1.5}$ assumed by \citet{sollins2005b} also implies an ionized-gas mass $0.5~M_\odot$, about $\times2$ larger than in our models.}, therefore the true value for $N_\mathrm{Ly}$ should be somewhere in between, although probably closer to our estimate. Using the stellar calibrations of \citet{martins2005}, the range $N_\mathrm{Ly} = 10^{49}$ to $10^{50}$ s$^{-1}$ is equivalent to the output of one Zero Age Main Sequence (ZAMS) star with mass $M_\star \approx 32$ to $\gtrsim 60~M_\odot$. The corresponding stellar luminosity is $L_\star \approx 2\times10^5$ to $ 1\times10^6~L_\odot$, which is well within the bolometric luminosity of the G10.6 region reported by \citet{lin2016}, $L_\mathrm{FIR} \approx 3\times10^6~L_\odot$.

The kinematical and luminosity constraints are satisfied by a number of stellar-mass arrangements, e.g., three stars of $50~M_\odot$ each provide enough gravity with a total $N_\mathrm{Ly} \approx 1\times 10^{50}$ s$^{-1}$. In the scenario of having $N_\mathrm{Ly}$ closer to our lower estimate $\sim 1\times 10^{49}$ s$^{-1}$, one star with $M_\star \approx 35~M_\odot$ would be enough to provide the ionizing photons, but the kinematical constraints would require the presence of an unknown number or lower-mass stars. Assuming that this unseen stellar cluster follows a \citet{kroupa2001} IMF, we calculate that the median cluster mass corresponding to a median maximum stellar mass $M_\mathrm{\star,max} = 35~M_\odot$ is $M_\mathrm{cl} \approx 500~M_\odot$, subject to significant stochasticity. We conclude that it is unlikely that a cluster sampling the full IMF is forming in such a reduced volume (see below), but an unseen group of lower-mass stars might coexist with the ionizing sources. Further observations at higher angular resolution from the radio to the mid-IR with the JWST are needed to clarify the distribution of (proto)stellar masses. 

For question $(ii)$, the current accretion stage depends on the interpretation of the ionized-gas kinematics. While the evidence for infall and rotation in the molecular gas beyond the central ionized structure is strong \citep{sollins2005b,liu2017}, our modelling shows that the radial motions in the ionized gas -- regardless of their direction -- become important starting somewhere in between $r \sim 1000$ to 1500 au, or about $r \sim R_g$. This suggests outward radial motions in the ionized gas.
This scenario would be similar to that of an ionized disk wind surrounded by collapsing molecular gas. A few cases of ionized disks have been reported around young massive stars  \citep[e.g.,][]{hoare2006,maud2018,guzman2020,jimenez2020}, but for objects that are less massive and luminous. Since the aforementioned objects are less embedded, their most likely interpretation is that of a remnant ionized disk without active accretion to the central star. The situation in G10.6 could be different. Even if the mass reservoir in the ionized disk is smaller than a solar mass, exterior to it exists a reservoir of infalling molecular gas that surpasses the central stellar mass beyond a radius of $\sim 0.1$ pc, and reaches up to $M_\mathrm{gas}\sim 2500$ M$_\odot$ at a radius of 0.5 pc \citep[see the mass profiles in][]{liu2017}. The molecular gas could replenish the ionized reservoir and the observed boundary of the ionized and molecular emission will depend on the asymmetric interactions of the ionizing photons and the surrounding molecular gas \citep{peters2010b,galvan2011}. Under the disk plus Ulrich envelope scenario, the  timescale for the replenishment of available $0.2~M_\odot$ of ionized gas is $t_\mathrm{rep} = M_\mathrm{ion-gas}/\dot{M}_\mathrm{env} \approx 6000$ yr, or $\times 10$ shorter than the expected star-formation timescale. Therefore, we conclude that even if replenishment of ionized gas occurs, it can provide at most a few $M_\odot$ of fresh material. Any active accretion ought to be in a residual stage. 

For question $(iii)$, the mass density of this young and compact cluster is large. Taking the upper and lower limits in mass ($M_\star = 120$ to $200~M_\odot$) and radius ($r_\star = 1000$ to 1500 au), we estimate $\rho_\star \sim 7.4\times10^7$ to $4.2\times10^8$ $M_\odot$ pc$^{-3}$. 
Constraining the stellar number density is challenging because of the aforementioned lack of knowledge of the lower-mass stellar population, but a number of stars in the range 5 to 20 translates
into $n_\star \sim 1\times10^7$ pc$^{-3}$. Our lower limit to $\rho_\star$ is slightly larger than the value derived by \citet{schoedel18} for the innermost 0.01 pc of the nuclear star cluster around Sgr A$^\ast$, $\rho_\mathrm{GC} = 2.6\times10^7$ $M_\odot$ pc$^{-3}$. However, the Galactic Center cluster is composed of $\sim 200$ solar-type stars. Our loose constraints on the number densities indicate that stellar interactions are likely to occur within the next Myr \citep{moeckel2007,zinnecker2007}.  

\section{Conclusions}

Using ALMA, we report the first kinematically resolved observations of an ionized rotating disk around a forming star cluster. The target is the luminous star formation region G10.6-0.4. We use radiative transfer models of the 1.3-mm free-free continuum and H$30\alpha$ line emission to constrain the density and velocity structure within a radius of 6000 au. Our preferred best-fit model is that of an ionized Keplerian disk with radial motions beyond a radius  $R_g \sim 1000$ to 1500 au, which corresponds to the radius within which the ionized gas is expected to be bound. The central stellar mass is robustly constrained to be in the range $M_\star = 120$ to $200~M_\odot$. The ionized-gas mass is only $M_\mathrm{ion-gas} = 0.2$ to $0.25~M_\odot$.    
The viewing inclination angle from face-on is in the range $i = 49$ to $56~\deg$. 
The fitted radial motions $v_{r,0} = 8.7$ \kms~correspond exactly to the sound speed ($c_s=8.6$ \kms) of ionized gas at $T_e=9000$ K,  indicating that the outer ionized gas is barely unbound at most. 
From constraints on the amount of ionizing photons and FIR luminosity, we conclude that there are either a few massive stars with $M_\star = 32$ to $60~M_\odot$, or one such massive star accompanied by an unknown number of lower-mass stars. 
Any active accretion of ionized gas onto the (proto)stars is mostly residual. 
The inferred cluster density is large, which suggests that stellar interactions are likely to occur within the next Myr .

\facilities{ALMA, SMA}

\software{
\texttt{CASA} \citep{mcmullin2007}, 
\texttt{Astropy} \citep{astropy18}, 
\texttt{sf3dmodels} \citep{izquierdo18}, 
\texttt{RADMC-3D} \citep{dullemond12}, 
\texttt{emcee} \citep{foreman13}, 
\texttt{corner} \citep{corner}, 
\texttt{spectral-cube} (\url{https://spectral-cube.readthedocs.io}), 
\texttt{pvextractor} (\url{https://pvextractor.readthedocs.io}), 
\texttt{IMF} (\url{https://github.com/keflavich/imf})
}

\acknowledgements 
This paper makes use of the following ALMA data: ADS/JAO.ALMA\#2015.1.00106.S. ALMA is a partnership of ESO (representing its member states), NSF (USA) and NINS (Japan), together with NRC (Canada), MOST and ASIAA (Taiwan), and KASI (Republic of Korea), in cooperation with the Republic of Chile. The Joint ALMA Observatory is operated by ESO, AUI/NRAO and NAOJ.
The National Radio Astronomy Observatory is a facility of the National Science Foundation operated under cooperative agreement by Associated Universities, Inc.
 R.G.-M. and C.C.-G. acknowledge support from UNAM-PAPIIT projects IN108822 and IG101321, and from CONACyT Ciencia de Frontera project ID: 86372. 
 R.G-M also acknowledges support from the AAS 
Chr\'etien International Research Grant. 
 H.B.L. is  supported by the National Science and Technology Council (NSTC) of Taiwan (Grant Nos. 108-2112-M-001-002-MY3, 111-2112-M-001-089-MY3, and 110-2112-M-001-069).
A.G. acknowledges support from NSF AAG 2008101 and NSF CAREER 2142300.

\bibliographystyle{aasjournal}

\bibliography{bibliography}

\begin{thebibliography}{}
\expandafter\ifx\csname natexlab\endcsname\relax\def\natexlab#1{#1}\fi
\providecommand{\url}[1]{\href{#1}{#1}}
\providecommand{\dodoi}[1]{doi:~\href{http://doi.org/#1}{\nolinkurl{#1}}}
\providecommand{\doeprint}[1]{\href{http://ascl.net/#1}{\nolinkurl{http://ascl.net/#1}}}
\providecommand{\doarXiv}[1]{\href{https://arxiv.org/abs/#1}{\nolinkurl{https://arxiv.org/abs/#1}}}

\bibitem[{{Astropy Collaboration} {et~al.}(2018){Astropy Collaboration},
  {Price-Whelan}, {Sip{\H{o}}cz}, {G{\"u}nther}, {Lim}, {Crawford}, {Conseil},
  {Shupe}, {Craig}, {Dencheva}, {Ginsburg}, {VanderPlas}, {Bradley},
  {P{\'e}rez-Su{\'a}rez}, {de Val-Borro}, {Aldcroft}, {Cruz}, {Robitaille},
  {Tollerud}, {Ardelean}, {Babej}, {Bach}, {Bachetti}, {Bakanov}, {Bamford},
  {Barentsen}, {Barmby}, {Baumbach}, {Berry}, {Biscani}, {Boquien}, {Bostroem},
  {Bouma}, {Brammer}, {Bray}, {Breytenbach}, {Buddelmeijer}, {Burke},
  {Calderone}, {Cano Rodr{\'\i}guez}, {Cara}, {Cardoso}, {Cheedella}, {Copin},
  {Corrales}, {Crichton}, {D'Avella}, {Deil}, {Depagne}, {Dietrich}, {Donath},
  {Droettboom}, {Earl}, {Erben}, {Fabbro}, {Ferreira}, {Finethy}, {Fox},
  {Garrison}, {Gibbons}, {Goldstein}, {Gommers}, {Greco}, {Greenfield},
  {Groener}, {Grollier}, {Hagen}, {Hirst}, {Homeier}, {Horton}, {Hosseinzadeh},
  {Hu}, {Hunkeler}, {Ivezi{\'c}}, {Jain}, {Jenness}, {Kanarek}, {Kendrew},
  {Kern}, {Kerzendorf}, {Khvalko}, {King}, {Kirkby}, {Kulkarni}, {Kumar},
  {Lee}, {Lenz}, {Littlefair}, {Ma}, {Macleod}, {Mastropietro}, {McCully},
  {Montagnac}, {Morris}, {Mueller}, {Mumford}, {Muna}, {Murphy}, {Nelson},
  {Nguyen}, {Ninan}, {N{\"o}the}, {Ogaz}, {Oh}, {Parejko}, {Parley}, {Pascual},
  {Patil}, {Patil}, {Plunkett}, {Prochaska}, {Rastogi}, {Reddy Janga},
  {Sabater}, {Sakurikar}, {Seifert}, {Sherbert}, {Sherwood-Taylor}, {Shih},
  {Sick}, {Silbiger}, {Singanamalla}, {Singer}, {Sladen}, {Sooley},
  {Sornarajah}, {Streicher}, {Teuben}, {Thomas}, {Tremblay}, {Turner},
  {Terr{\'o}n}, {van Kerkwijk}, {de la Vega}, {Watkins}, {Weaver}, {Whitmore},
  {Woillez}, {Zabalza}, \& {Astropy Contributors}}]{astropy18}
{Astropy Collaboration}, {Price-Whelan}, A.~M., {Sip{\H{o}}cz}, B.~M., {et~al.}
  2018, \aj, 156, 123, \dodoi{10.3847/1538-3881/aabc4f}

\bibitem[{{Beltr{\'a}n} {et~al.}(2011){Beltr{\'a}n}, {Cesaroni}, {Neri}, \&
  {Codella}}]{beltran2011}
{Beltr{\'a}n}, M.~T., {Cesaroni}, R., {Neri}, R., \& {Codella}, C. 2011, \aap,
  525, A151, \dodoi{10.1051/0004-6361/201015049}

\bibitem[{{Dullemond} {et~al.}(2012){Dullemond}, {Juhasz}, {Pohl}, {Sereshti},
  {Shetty}, {Peters}, {Commercon}, \& {Flock}}]{dullemond12}
{Dullemond}, C.~P., {Juhasz}, A., {Pohl}, A., {et~al.} 2012, {RADMC-3D: A
  multi-purpose radiative transfer tool}, Astrophysics Source Code Library,
  record ascl:1202.015.
\newblock \doeprint{1202.015}

\bibitem[{Foreman-Mackey(2016)}]{corner}
Foreman-Mackey, D. 2016, The Journal of Open Source Software, 1, 24,
  \dodoi{10.21105/joss.00024}

\bibitem[{{Foreman-Mackey} {et~al.}(2013){Foreman-Mackey}, {Hogg}, {Lang}, \&
  {Goodman}}]{foreman13}
{Foreman-Mackey}, D., {Hogg}, D.~W., {Lang}, D., \& {Goodman}, J. 2013, \pasp,
  125, 306, \dodoi{10.1086/670067}

\bibitem[{{Franco} {et~al.}(1989){Franco}, {Tenorio-Tagle}, \&
  {Bodenheimer}}]{franco1989}
{Franco}, J., {Tenorio-Tagle}, G., \& {Bodenheimer}, P. 1989, \rmxaa, 18, 65

\bibitem[{{Galv{\'a}n-Madrid} {et~al.}(2011){Galv{\'a}n-Madrid}, {Peters},
  {Keto}, {Mac Low}, {Banerjee}, \& {Klessen}}]{galvan2011}
{Galv{\'a}n-Madrid}, R., {Peters}, T., {Keto}, E.~R., {et~al.} 2011, \mnras,
  416, 1033, \dodoi{10.1111/j.1365-2966.2011.19101.x}

\bibitem[{{Giacobbo} \& {Mapelli}(2018)}]{giacobbo2018}
{Giacobbo}, N., \& {Mapelli}, M. 2018, \mnras, 480, 2011,
  \dodoi{10.1093/mnras/sty1999}

\bibitem[{{Goddi} {et~al.}(2020){Goddi}, {Ginsburg}, {Maud}, {Zhang}, \&
  {Zapata}}]{goddi2020}
{Goddi}, C., {Ginsburg}, A., {Maud}, L.~T., {Zhang}, Q., \& {Zapata}, L.~A.
  2020, \apj, 905, 25, \dodoi{10.3847/1538-4357/abc88e}

\bibitem[{{Guzm{\'a}n} {et~al.}(2020){Guzm{\'a}n}, {Sanhueza}, {Zapata},
  {Garay}, \& {Rodr{\'\i}guez}}]{guzman2020}
{Guzm{\'a}n}, A.~E., {Sanhueza}, P., {Zapata}, L., {Garay}, G., \&
  {Rodr{\'\i}guez}, L.~F. 2020, \apj, 904, 77, \dodoi{10.3847/1538-4357/abbe09}

\bibitem[{{Ho} \& {Haschick}(1986)}]{ho1986}
{Ho}, P.~T.~P., \& {Haschick}, A.~D. 1986, \apj, 304, 501,
  \dodoi{10.1086/164184}

\bibitem[{{Hoare}(2006)}]{hoare2006}
{Hoare}, M.~G. 2006, \apj, 649, 856, \dodoi{10.1086/506961}

\bibitem[{{Hollenbach} {et~al.}(1994){Hollenbach}, {Johnstone}, {Lizano}, \&
  {Shu}}]{hollenbach1994}
{Hollenbach}, D., {Johnstone}, D., {Lizano}, S., \& {Shu}, F. 1994, \apj, 428,
  654

\bibitem[{{Hosokawa} {et~al.}(2010){Hosokawa}, {Yorke}, \&
  {Omukai}}]{hosokawa10}
{Hosokawa}, T., {Yorke}, H.~W., \& {Omukai}, K. 2010, \apj, 721, 478,
  \dodoi{10.1088/0004-637X/721/1/478}

\bibitem[{{Ilee} {et~al.}(2018){Ilee}, {Cyganowski}, {Brogan}, {Hunter},
  {Forgan}, {Haworth}, {Clarke}, \& {Harries}}]{ilee2018}
{Ilee}, J.~D., {Cyganowski}, C.~J., {Brogan}, C.~L., {et~al.} 2018, \apjl, 869,
  L24, \dodoi{10.3847/2041-8213/aaeffc}

\bibitem[{{Izquierdo} {et~al.}(2018){Izquierdo}, {Galv{\'a}n-Madrid}, {Maud},
  {Hoare}, {Johnston}, {Keto}, {Zhang}, \& {de Wit}}]{izquierdo18}
{Izquierdo}, A.~F., {Galv{\'a}n-Madrid}, R., {Maud}, L.~T., {et~al.} 2018,
  \mnras, 478, 2505, \dodoi{10.1093/mnras/sty1096}

\bibitem[{{Jim{\'e}nez-Serra} {et~al.}(2020){Jim{\'e}nez-Serra},
  {B{\'a}ez-Rubio}, {Mart{\'\i}n-Pintado}, {Zhang}, \& {Rivilla}}]{jimenez2020}
{Jim{\'e}nez-Serra}, I., {B{\'a}ez-Rubio}, A., {Mart{\'\i}n-Pintado}, J.,
  {Zhang}, Q., \& {Rivilla}, V.~M. 2020, \apjl, 897, L33,
  \dodoi{10.3847/2041-8213/aba050}

\bibitem[{{Johnston} {et~al.}(2015){Johnston}, {Robitaille}, {Beuther}, {Linz},
  {Boley}, {Kuiper}, {Keto}, {Hoare}, \& {van Boekel}}]{johnston2015}
{Johnston}, K.~G., {Robitaille}, T.~P., {Beuther}, H., {et~al.} 2015, \apjl,
  813, L19, \dodoi{10.1088/2041-8205/813/1/L19}

\bibitem[{{Keto}(2002)}]{keto2002b}
{Keto}, E. 2002, \apj, 580, 980

\bibitem[{{Keto}(2003)}]{keto2003}
---. 2003, \apj, 599, 1196

\bibitem[{{Keto}(2007)}]{keto2007}
---. 2007, \apj, 666, 976, \dodoi{10.1086/520320}

\bibitem[{{Keto} \& {Wood}(2006)}]{keto2006}
{Keto}, E., \& {Wood}, K. 2006, \apj, 637, 850, \dodoi{10.1086/498611}

\bibitem[{{Keto} \& {Zhang}(2010)}]{keto2010}
{Keto}, E., \& {Zhang}, Q. 2010, \mnras, 406, 102,
  \dodoi{10.1111/j.1365-2966.2010.16672.x}

\bibitem[{{Keto} {et~al.}(1987){Keto}, {Ho}, \& {Haschick}}]{keto1987a}
{Keto}, E.~R., {Ho}, P.~T.~P., \& {Haschick}, A.~D. 1987, \apj, 318, 712,
  \dodoi{10.1086/165405}

\bibitem[{{Klaassen} {et~al.}(2018){Klaassen}, {Johnston}, {Urquhart},
  {Mottram}, {Peters}, {Kuiper}, {Beuther}, {van der Tak}, \&
  {Goddi}}]{klaassen2018}
{Klaassen}, P.~D., {Johnston}, K.~G., {Urquhart}, J.~S., {et~al.} 2018, \aap,
  611, A99, \dodoi{10.1051/0004-6361/201731727}

\bibitem[{{Kroupa}(2001)}]{kroupa2001}
{Kroupa}, P. 2001, \mnras, 322, 231, \dodoi{10.1046/j.1365-8711.2001.04022.x}

\bibitem[{{Krumholz} {et~al.}(2014){Krumholz}, {Bate}, {Arce}, {Dale},
  {Gutermuth}, {Klein}, {Li}, {Nakamura}, \& {Zhang}}]{krumholz2014}
{Krumholz}, M.~R., {Bate}, M.~R., {Arce}, H.~G., {et~al.} 2014, Protostars and
  Planets VI, 243, \dodoi{10.2458/azu_uapress_9780816531240-ch011}

\bibitem[{{Law} {et~al.}(2021){Law}, {Zhang}, {{\"O}berg}, {Galv{\'a}n-Madrid},
  {Keto}, {Liu}, \& {Ho}}]{law21}
{Law}, C.~J., {Zhang}, Q., {{\"O}berg}, K.~I., {et~al.} 2021, \apj, 909, 214,
  \dodoi{10.3847/1538-4357/abdeb8}

\bibitem[{{Lin} {et~al.}(2016){Lin}, {Liu}, {Li}, {Zhang}, {Ginsburg},
  {Pineda}, {Qian}, {Galv{\'a}n-Madrid}, {McLeod}, {Rosolowsky}, {Dale},
  {Immer}, {Koch}, {Longmore}, {Walker}, \& {Testi}}]{lin2016}
{Lin}, Y., {Liu}, H.~B., {Li}, D., {et~al.} 2016, \apj, 828, 32,
  \dodoi{10.3847/0004-637X/828/1/32}

\bibitem[{{Liu} {et~al.}(2010){Liu}, {Ho}, {Zhang}, {Keto}, {Wu}, \&
  {Li}}]{liu2010b}
{Liu}, H., {Ho}, P.~T.~P., {Zhang}, Q., {et~al.} 2010, \apj, 722, 262,
  \dodoi{10.1088/0004-637X/722/1/262}

\bibitem[{{Liu}(2017)}]{liu2017}
{Liu}, H.~B. 2017, \aap, 597, A70, \dodoi{10.1051/0004-6361/201629582}

\bibitem[{{Liu} {et~al.}(2011){Liu}, {Zhang}, \& {Ho}}]{liu2011}
{Liu}, H.~B., {Zhang}, Q., \& {Ho}, P.~T.~P. 2011, \apj, 729, 100,
  \dodoi{10.1088/0004-637X/729/2/100}

\bibitem[{{Lu} {et~al.}(2022){Lu}, {Li}, {Zhang}, \& {Lin}}]{lu2022}
{Lu}, X., {Li}, G.-X., {Zhang}, Q., \& {Lin}, Y. 2022, Nature Astronomy, 6,
  837, \dodoi{10.1038/s41550-022-01681-4}

\bibitem[{{Martins} {et~al.}(2005){Martins}, {Schaerer}, \&
  {Hillier}}]{martins2005}
{Martins}, F., {Schaerer}, D., \& {Hillier}, D.~J. 2005, \aap, 436, 1049,
  \dodoi{10.1051/0004-6361:20042386}

\bibitem[{{Maud} {et~al.}(2018){Maud}, {Cesaroni}, {Kumar}, {van der Tak},
  {Allen}, {Hoare}, {Klaassen}, {Harsono}, {Hogerheijde}, {S{\'a}nchez-Monge},
  {Schilke}, {Ahmadi}, {Beltr{\'a}n}, {Beuther}, {Csengeri}, {Etoka}, {Fuller},
  {Galv{\'a}n-Madrid}, {Goddi}, {Henning}, {Johnston}, {Kuiper}, {Lumsden},
  {Moscadelli}, {Mottram}, {Peters}, {Rivilla}, {Testi}, {Vig}, {de Wit}, \&
  {Zinnecker}}]{maud2018}
{Maud}, L.~T., {Cesaroni}, R., {Kumar}, M.~S.~N., {et~al.} 2018, \aap, 620,
  A31, \dodoi{10.1051/0004-6361/201833908}

\bibitem[{{McMullin} {et~al.}(2007){McMullin}, {Waters}, {Schiebel}, {Young},
  \& {Golap}}]{mcmullin2007}
{McMullin}, J.~P., {Waters}, B., {Schiebel}, D., {Young}, W., \& {Golap}, K.
  2007, in Astronomical Society of the Pacific Conference Series, Vol. 376,
  Astronomical Data Analysis Software and Systems XVI, ed. R.~A. {Shaw},
  F.~{Hill}, \& D.~J. {Bell}, 127

\bibitem[{{Mendoza} {et~al.}(2004){Mendoza}, {Cant{\'o}}, \&
  {Raga}}]{mendoza04}
{Mendoza}, S., {Cant{\'o}}, J., \& {Raga}, A.~C. 2004, \rmxaa, 40, 147.
\newblock \doarXiv{astro-ph/0401426}

\bibitem[{{Moeckel} \& {Bally}(2007)}]{moeckel2007}
{Moeckel}, N., \& {Bally}, J. 2007, \apjl, 661, L183, \dodoi{10.1086/518738}

\bibitem[{{Olguin} {et~al.}(2020){Olguin}, {Hoare}, {Johnston}, {Motte},
  {Chen}, {Beuther}, {Mottram}, {Ahmadi}, {Gieser}, {Semenov}, {Peters},
  {Palau}, {Klaassen}, {Kuiper}, {S{\'a}nchez-Monge}, \&
  {Henning}}]{olguin2020}
{Olguin}, F.~A., {Hoare}, M.~G., {Johnston}, K.~G., {et~al.} 2020, \mnras, 498,
  4721, \dodoi{10.1093/mnras/staa2406}

\bibitem[{{Palla} \& {Stahler}(1993)}]{palla1993}
{Palla}, F., \& {Stahler}, S.~W. 1993, \apj, 418, 414

\bibitem[{{Peters} {et~al.}(2010{\natexlab{a}}){Peters}, {Banerjee}, {Klessen},
  {Mac Low}, {Galv{\'a}n-Madrid}, \& {Keto}}]{peters2010b}
{Peters}, T., {Banerjee}, R., {Klessen}, R.~S., {et~al.} 2010{\natexlab{a}},
  \apj, 711, 1017, \dodoi{10.1088/0004-637X/711/2/1017}

\bibitem[{{Peters} {et~al.}(2012){Peters}, {Longmore}, \&
  {Dullemond}}]{peters2012}
{Peters}, T., {Longmore}, S.~N., \& {Dullemond}, C.~P. 2012, \mnras, 425, 2352,
  \dodoi{10.1111/j.1365-2966.2012.21676.x}

\bibitem[{{Peters} {et~al.}(2010{\natexlab{b}}){Peters}, {Mac Low}, {Banerjee},
  {Klessen}, \& {Dullemond}}]{peters2010a}
{Peters}, T., {Mac Low}, M.-M., {Banerjee}, R., {Klessen}, R.~S., \&
  {Dullemond}, C.~P. 2010{\natexlab{b}}, \apj, 719, 831,
  \dodoi{10.1088/0004-637X/719/1/831}

\bibitem[{{Pringle}(1981)}]{pringle1981}
{Pringle}, J.~E. 1981, \araa, 19, 137,
  \dodoi{10.1146/annurev.aa.19.090181.001033}

\bibitem[{{Rohlfs} \& {Wilson}(2000)}]{rohlfs2000}
{Rohlfs}, K., \& {Wilson}, T.~L. 2000, {Tools of radio astronomy} (Tools of
  radio astronomy / K.~Rohlfs, T.L.~Wilson.~New York : Springer,
  2000.~(Astronomy and astrophysics library,ISSN0941-7834))

\bibitem[{{Sanna} {et~al.}(2014){Sanna}, {Reid}, {Menten}, {Dame}, {Zhang},
  {Sato}, {Brunthaler}, {Moscadelli}, \& {Immer}}]{sanna2014}
{Sanna}, A., {Reid}, M.~J., {Menten}, K.~M., {et~al.} 2014, \apj, 781, 108,
  \dodoi{10.1088/0004-637X/781/2/108}

\bibitem[{{Sanna} {et~al.}(2019){Sanna}, {K{\"o}lligan}, {Moscadelli},
  {Kuiper}, {Cesaroni}, {Pillai}, {Menten}, {Zhang}, {Caratti o Garatti},
  {Goddi}, {Leurini}, \& {Carrasco-Gonz{\'a}lez}}]{sanna2019}
{Sanna}, A., {K{\"o}lligan}, A., {Moscadelli}, L., {et~al.} 2019, \aap, 623,
  A77, \dodoi{10.1051/0004-6361/201833411}

\bibitem[{{Sch{\"o}del} {et~al.}(2018){Sch{\"o}del}, {Gallego-Cano}, {Dong},
  {Nogueras-Lara}, {Gallego-Calvente}, {Amaro-Seoane}, \&
  {Baumgardt}}]{schoedel18}
{Sch{\"o}del}, R., {Gallego-Cano}, E., {Dong}, H., {et~al.} 2018, \aap, 609,
  A27, \dodoi{10.1051/0004-6361/201730452}

\bibitem[{{Sollins} {et~al.}(2005){Sollins}, {Zhang}, {Keto}, \&
  {Ho}}]{sollins2005b}
{Sollins}, P.~K., {Zhang}, Q., {Keto}, E., \& {Ho}, P.~T.~P. 2005, \apj, 631,
  399, \dodoi{10.1086/432503}

\bibitem[{{Tanaka} {et~al.}(2013){Tanaka}, {Nakamoto}, \&
  {Omukai}}]{tanaka2013}
{Tanaka}, K. E.~I., {Nakamoto}, T., \& {Omukai}, K. 2013, \apj, 773, 155,
  \dodoi{10.1088/0004-637X/773/2/155}

\bibitem[{{Tanaka} {et~al.}(2017){Tanaka}, {Tan}, \& {Zhang}}]{tanaka2017}
{Tanaka}, K. E.~I., {Tan}, J.~C., \& {Zhang}, Y. 2017, \apj, 835, 32,
  \dodoi{10.3847/1538-4357/835/1/32}

\bibitem[{{Ulrich}(1976)}]{ulrich1976}
{Ulrich}, R.~K. 1976, \apj, 210, 377

\bibitem[{{Zinnecker} \& {Yorke}(2007)}]{zinnecker2007}
{Zinnecker}, H., \& {Yorke}, H.~W. 2007, \araa, 45, 481,
  \dodoi{10.1146/annurev.astro.44.051905.092549}

\end{thebibliography}

\appendix 

\section{Diagnostic plots} \label{secc:app-diag}

\begin{figure}
\hspace{-2.0cm}
\centering
\includegraphics[width=\columnwidth]{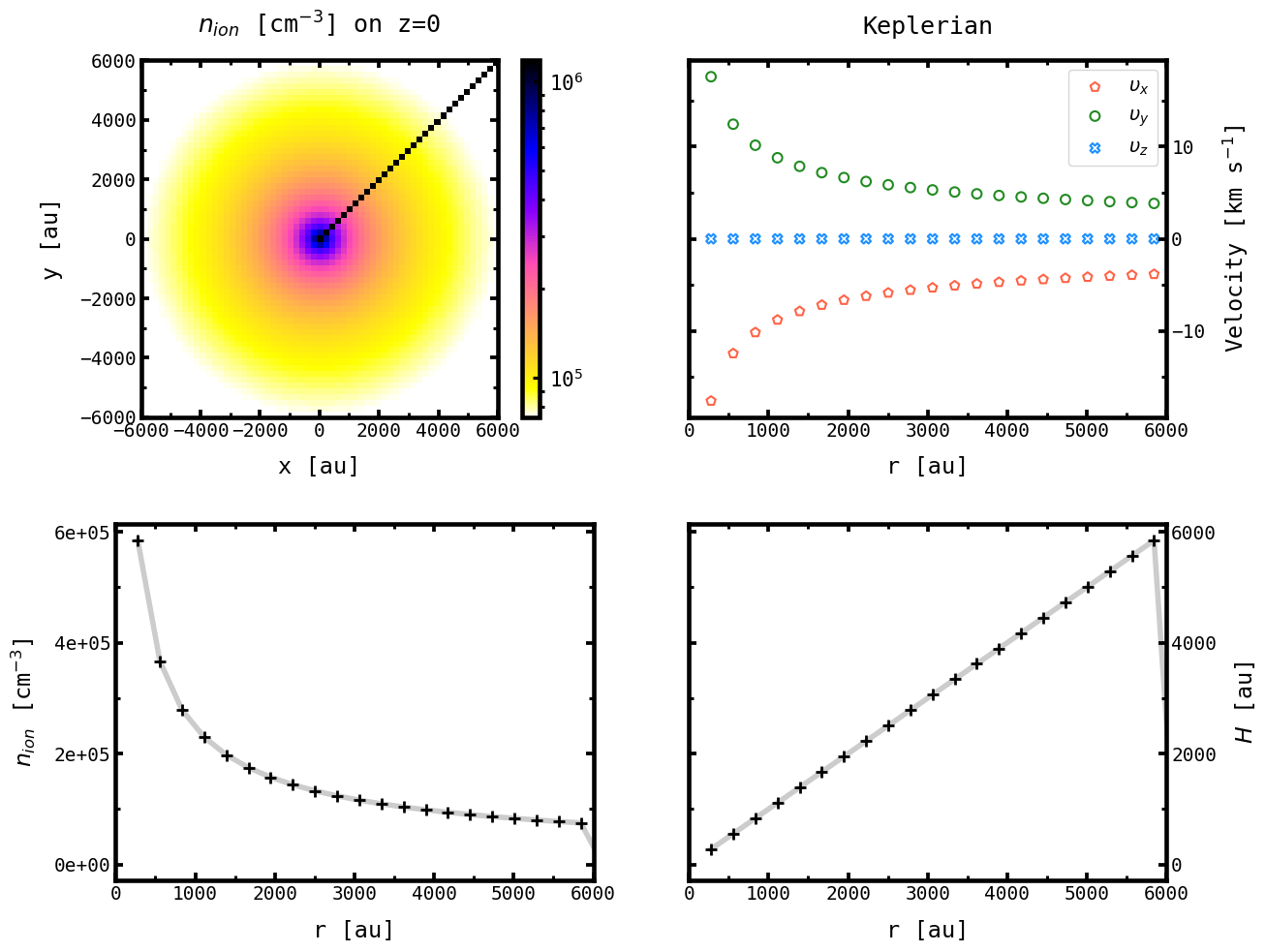}
\caption{Diagnostic plots for best fit of purely Keplerian model. The {\it top-left} panel shows the midplane ion density $n_{ion} = n_e$, along with a cut in which the velocity field is measured. The {\it top-right} panel shows cartesian velocity components in the midplane $z=0$ cut. The {\it bottom-left} panel shows the midplane density profile. The {\it bottom-right} panel shows the scale-height $H$ profile. 
}
\label{fig:App1-Fig1}
\end{figure}

\begin{figure*}
\hspace{-2.0cm}
\centering
\includegraphics[width=\textwidth]{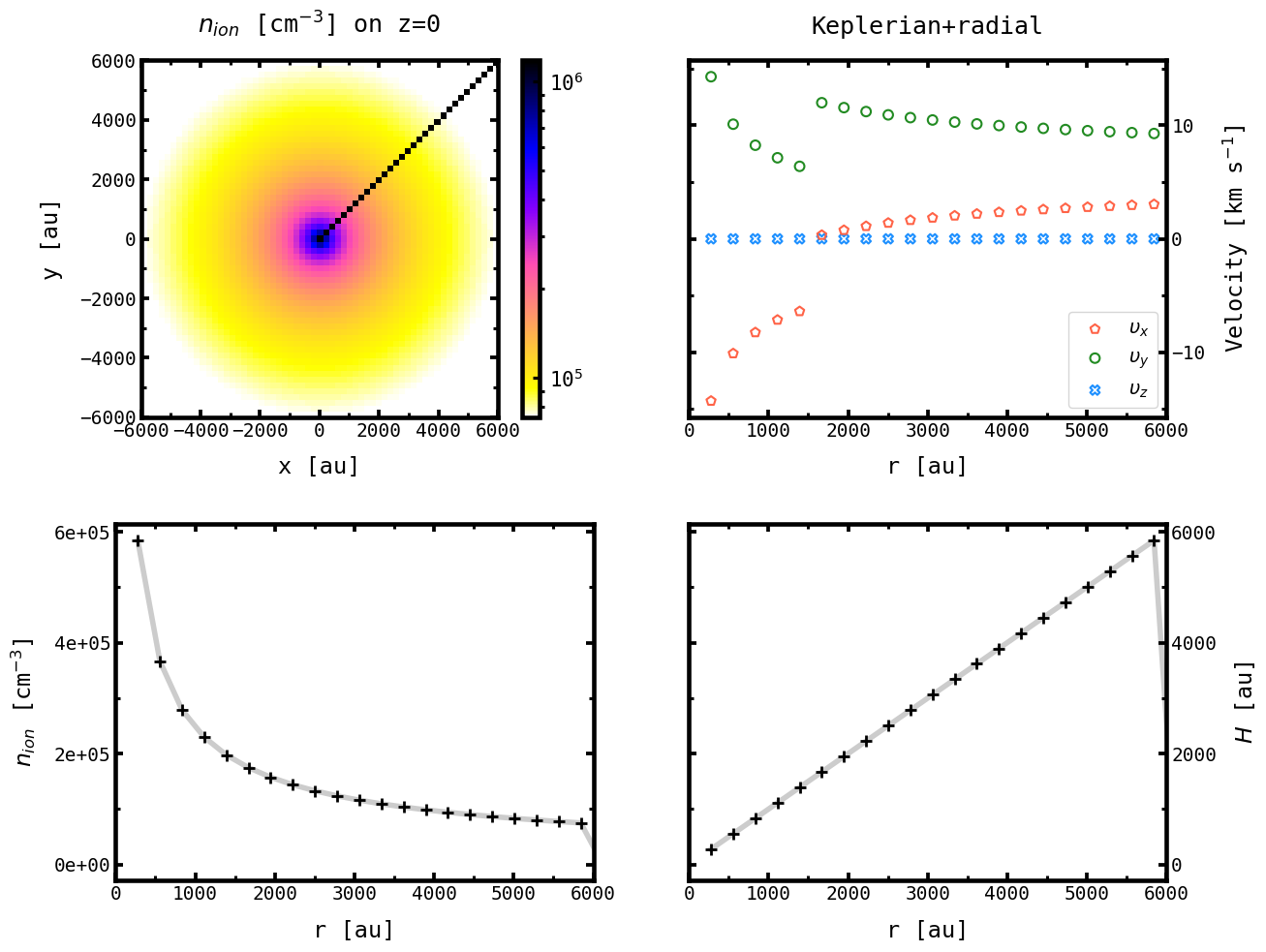}
\caption{Same as Fig. \ref{fig:App1-Fig1}, but for the case of the case of a Keplerian disk with external radial motions.}
\label{fig:App1-Fig2}
\end{figure*}

\begin{figure*}
\hspace{-2.0cm}
\centering
\includegraphics[width=\textwidth]{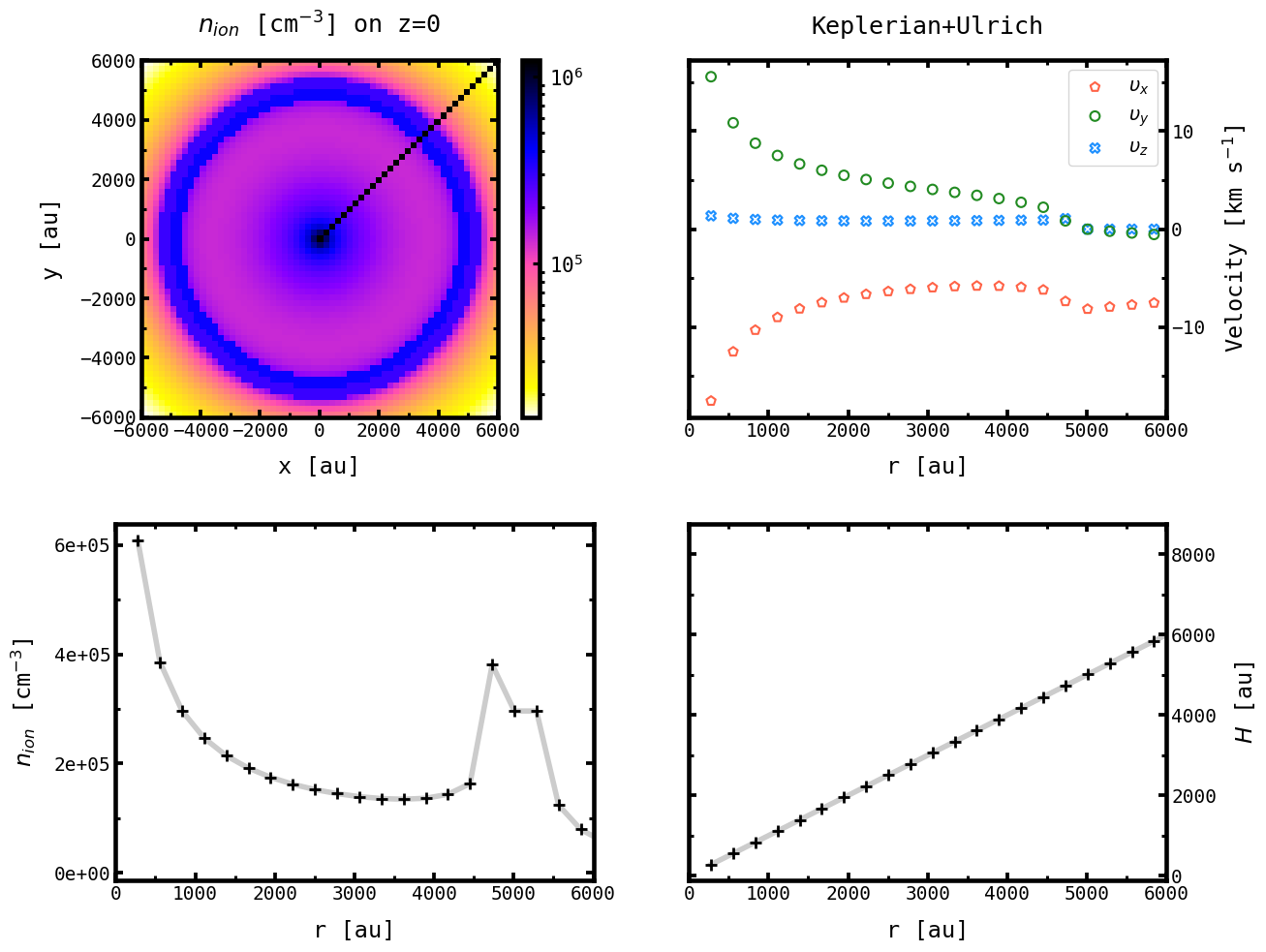}
\caption{Same as Fig. \ref{fig:App1-Fig1}, but for the case of the case of a Keplerian disk with an exterior Ulrich envelope. Note that the Ulrich envelope has a divergence at the radius in which it settles into a disk \citep{ulrich1976,mendoza04}, which we smooth by averaging a few neighboring grid points.}
\label{fig:App1-Fig3}
\end{figure*}

In this Appendix we show diagnostic plots for the best-fit models of each of the considered scenarios: purely Keplerian disk (Fig. \ref{fig:App1-Fig1}), Keplerian disk with radial motions starting at $r_0$ (Fig. \ref{fig:App1-Fig2}), and Keplerian disk with an external Ulrich-type envelope (Fig. \ref{fig:App1-Fig3}). 

\end{document}